\documentclass[11pt,a4paper]{elsarticle}

\usepackage{epsfig}

\usepackage{latexsym}
\usepackage{amsmath}
\usepackage{amsfonts}
\usepackage{amssymb}

\newcommand{\be}{\begin{equation}}
\newcommand{\ee}{\end{equation}}
\newcommand{\bea}{\begin{eqnarray}}
\newcommand{\eea}{\end{eqnarray}}
\newcommand{\Tr}{{\rm Tr}}

\journal{Physics Letters B}

\begin{document}

\begin{frontmatter}

\title{Large-N reduction of SU($N$) Yang-Mills theory with massive adjoint overlap fermions}
\author[label1,label2]{A. Hietanen} 
\ead{a.j.hietanen@swansea.ac.uk}
\author[label1]{R. Narayanan}
\ead{rajamani.narayanan@fiu.edu}
\address[label1]{Department of Physics, Florida International University, Miami,
FL 33199, USA}
\address[label2]{
School of Physical Sciences, Swansea University Singleton Park, Swansea SA2 8PP, UK
}

\begin{abstract}

We study four dimensional large-N SU$(N)$ Yang-Mills theory 
coupled to adjoint overlap fermions on a single site lattice. 
Lattice simulations along with perturbation theory show that the bare
quark mass has to be taken to zero as one takes the continuum limit in order
to be in the physically relevant center-symmetric phase. But, it seems that it is possible to
take the continuum limit with any renormalized quark mass and still
be in the center-symmetric physics.
We have also conducted a study of the correlations
between Polyakov loop operators in different directions and obtained the
range for the Wilson mass parameter that enters the overlap Dirac
operator.

\end{abstract}

\begin{keyword}
1/N Expansion \sep Eguchi-Kawai reduction \sep Adjoint fermions \sep
Lattice Gauge Field Theories
\end{keyword}

\end{frontmatter}

The large $N$ limit of gauge theories has many intriguing
properties. One of these is continuum
reduction~\cite{Narayanan:2003fc}. It states that one obtains correct
infinite volume zero temperature results by working on a finite volume
lattice as long as the center symmetry is
intact. In~\cite{Kovtun:2007py}, it was proposed, that for a
Yang-Mills theory with massless adjoint fermions with periodic
boundary conditions, the volume can be reduced down to a single
site as opposed to the pure gauge
case~\cite{Eguchi:1982nm}, where weak coupling analysis shows all the
center symmetries to be broken~\cite{Bhanot:1982sh}. This has been
confirmed both by lattice techniques and by perturbation theory
\cite{Hollowood:2009sy,Bringoltz:2009mi,Bringoltz:2009fj,Poppitz:2009fm,
Hietanen:2009ex,Bringoltz:2009kb,Hietanen:2010bz,Poppitz:2008hr,Catterall:2010gx}.

The question we want to address in this paper is what occurs at the large
$N$ continuum limit when fermions have a mass. The large $N$ continuum
limit is taken by first extrapolating $N\rightarrow\infty$ and then
$b\rightarrow\infty$, where $b$ is the inverse 't Hooft coupling,
$\frac{1}{g^2N}$. It has been argued in~\cite{Azeyanagi:2010ne} that for
any finite mass, a center symmetry unbroken phase exists at sufficiently
small volume. Lattice studies using Wilson fermions
has shown a  large range of masses at fixed lattice spacing where the
center symmetry
remains intact~\cite{Bringoltz:2009kb}.

In this paper, we address the question of center symmetry both in the
lattice
and in the continuum using massive adjoint overlap 
fermions~\cite{Neuberger:1997fp,Edwards:1998wx}.
We show that the critical bare quark mass $\mu_c$,
above which the center symmetry is broken, is zero at the continuum
limit. However, on a lattice with a finite lattice spacing, $\mu_c>0$.
Values of masses, which are accessible to lattice simulations, depend
on how $\mu_c$ scales as a function of the lattice spacing.

We study the problem with one Weyl fermion, $f=0.5$, both by
perturbation theory and lattice simulations. 
Using perturbation
theory we show that center symmetry is broken even
when quarks are given an arbitrarily small mass.
We have performed lattice simulations with different $b$ and
$N$. The lattice results confirm with perturbation theory and we find
a $\mu_c(b)$ that decreases as $b$ increases. 
We do not see any evidence of scaling of $\mu_c(b)$ versus $b$. Our
numerical
results indicate that we can obtain the continuum limit with
arbitrary physical mass for the adjoint quarks.

All details pertaining to the single site lattice model with adjoint
overlap fermions are described in~\cite{Hietanen:2009ex}.
To study the continuum limit starting from the single site action,
we use the weak coupling expansion and write the link matrices as
\be 
U_\mu = e^{ia_\mu} D_\mu e^{-ia_\mu};\ \ \ \ 
D_\mu^{ij}=e^{i\theta_\mu^i}\delta_{ij},
\ee
and perform an expansion in $a_\mu$. The $\theta_\mu^i$
are the eigenvalues of the Polyakov loop operator and they have to
uniformly distributed in the range $[-\pi,\pi]$ and uncorrelated in
all four directions in order to correctly reproduce infinite volume
continuum perturbation theory.
The leading order result is 
\begin{align}
&S =  \sum_{i\ne j}  \Bigg\{
\ln \left[\sum_\mu \sin^2 \frac{1}{2}\left(\theta_\mu^i-\theta_\mu^j\right) \right]
 \nonumber \\
&\left.-2f\ln \left[
\frac{1+\mu^2}{2} + \frac{1-\mu^2}{2} 
\frac{2\sum_\mu\sin^2\frac{\theta_\mu^i-\theta_\mu^j}{2} -m_w}
{\sqrt{
\left(2\sum_\mu\sin^2\frac{\theta_\mu^i-\theta_\mu^j}{2} -m_w\right)^2 + \sum_\mu \sin^2(\theta_\mu^i-\theta_\mu^j)}}\right]
\right\},
\label{pertact}
\end{align}
where the first line of RHS is the contribution from gauge
fields~\cite{Bhanot:1982sh} and the second line is the contribution
from $f$ flavors of Dirac fermions~\cite{Hietanen:2009ex}. The bare quark mass is
$\frac{2m_w\mu}{\sqrt{1-\mu^2}}$ with $\mu \in [0,1]$ and
$m_w$ is the Wilson mass parameter. 

The gauge action has its minimum, $-\infty$, when all the angles
$\theta_\mu^i$ are equal. With one massless Weyl fermion
($f=0.5$) the fermionic part cancels out the infinity and renders the
action finite. In~\cite{Hietanen:2009ex} we used Monte Carlo
techniques to find out the actual minimum. Namely, we consider the
Hamiltonian \be H = \frac{1}{2}\sum_{\mu,i} \left(\pi_\mu^i\right)^2 +
\beta S.  \ee For large $\beta$, the Boltzmann measure $e^{-H}$ is
dominated by the minimum. Hence, this minimum can be found by
performing a HMC update for the $\pi,\theta$ system.

To reduce rounding errors in equations of motions, we introduce a
regulator $\Delta$ to the gauge field action 
\be S_g \rightarrow
\sum_{i\ne j} \ln \left[\sum_\mu \sin^2
  \frac{1}{2}\left(\theta_\mu^i-\theta_\mu^j\right) +\Delta\right].
\ee 
In the computations we choose $\Delta=10^{-4}$, which is much
smaller than the average difference between angles $2\pi/N$ when
$N<200$.

A choice for the order parameters associated with the $Z_N^4$ symmetries is~\cite{Bhanot:1982sh}
\be
P_\mu = \frac{1}{2} \left( 1 - \frac{1}{N^2}|\Tr U_\mu|^2\right)
=\frac{1}{N^2}\sum_{i,j}
\sin^2 \frac{1}{2}\left(\theta_\mu^i-
\theta_\mu^j\right).
\ee
If $P_\mu=\frac12$, then the $Z_N$ symmetry in that direction is unbroken.

\begin{figure}
\centerline{\includegraphics[width=1.0\textwidth]{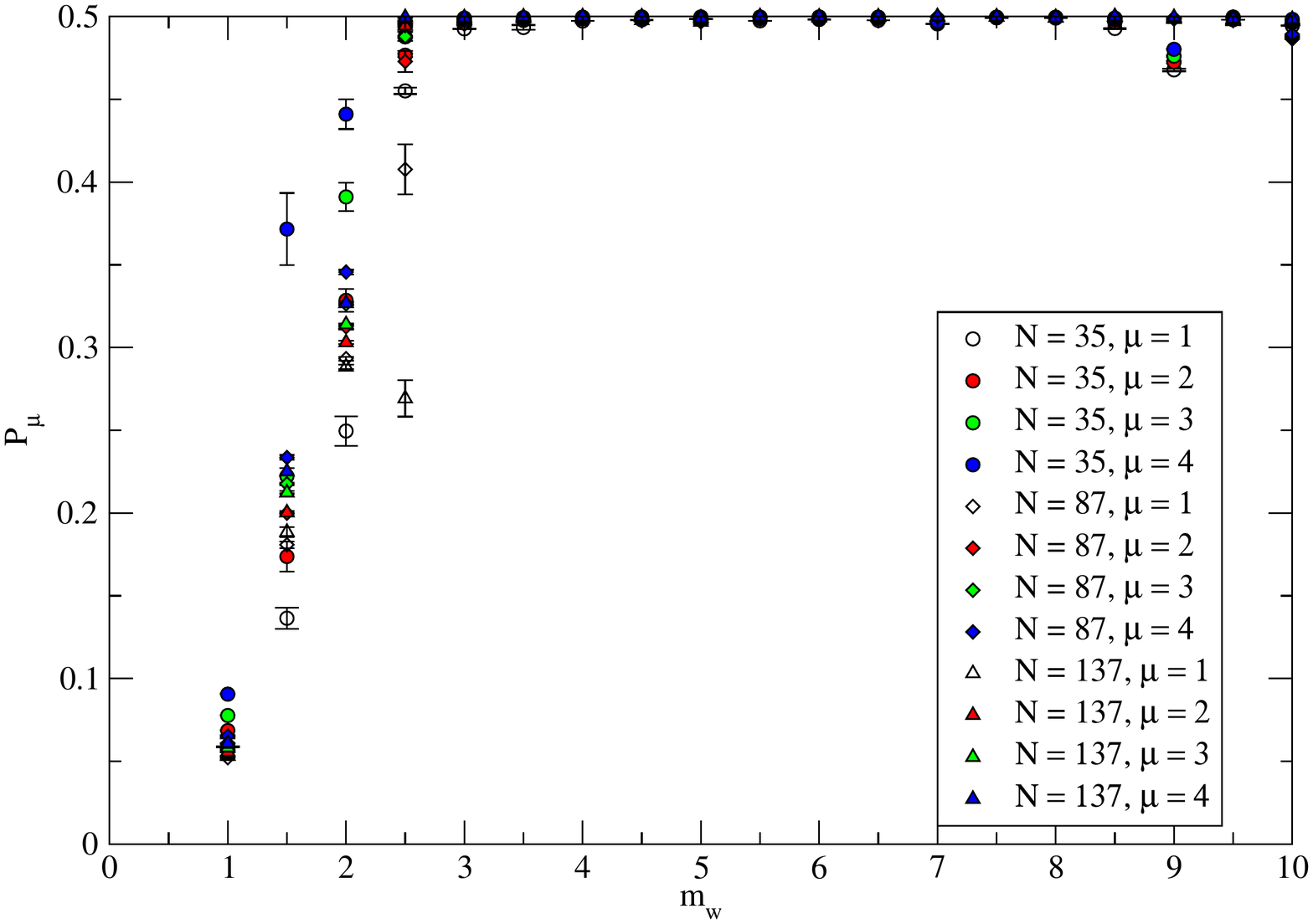}}
\caption{Plot of $P_\mu$ as a function of Wilson mass. \label{figSymmetry_m0}}
\end{figure}

In Fig.~\ref{figSymmetry_m0} we reproduce the results of
\cite{Hietanen:2009ex} with a large $N$ scaling. To better observe the
center symmetry breaking, the measurements $P_i$ are ordered for each
configuration s.t. $P_1<P_2<P_3<P_4$.  This indicates that center
symmetry is probably restored when Wilson mass is in the range $3.0< m_w < 10.0$. The range of $m_w$ with broken symmetry does not depend on $N$.

\begin{figure}
  \begin{center}
    \includegraphics[width=1.0\textwidth]{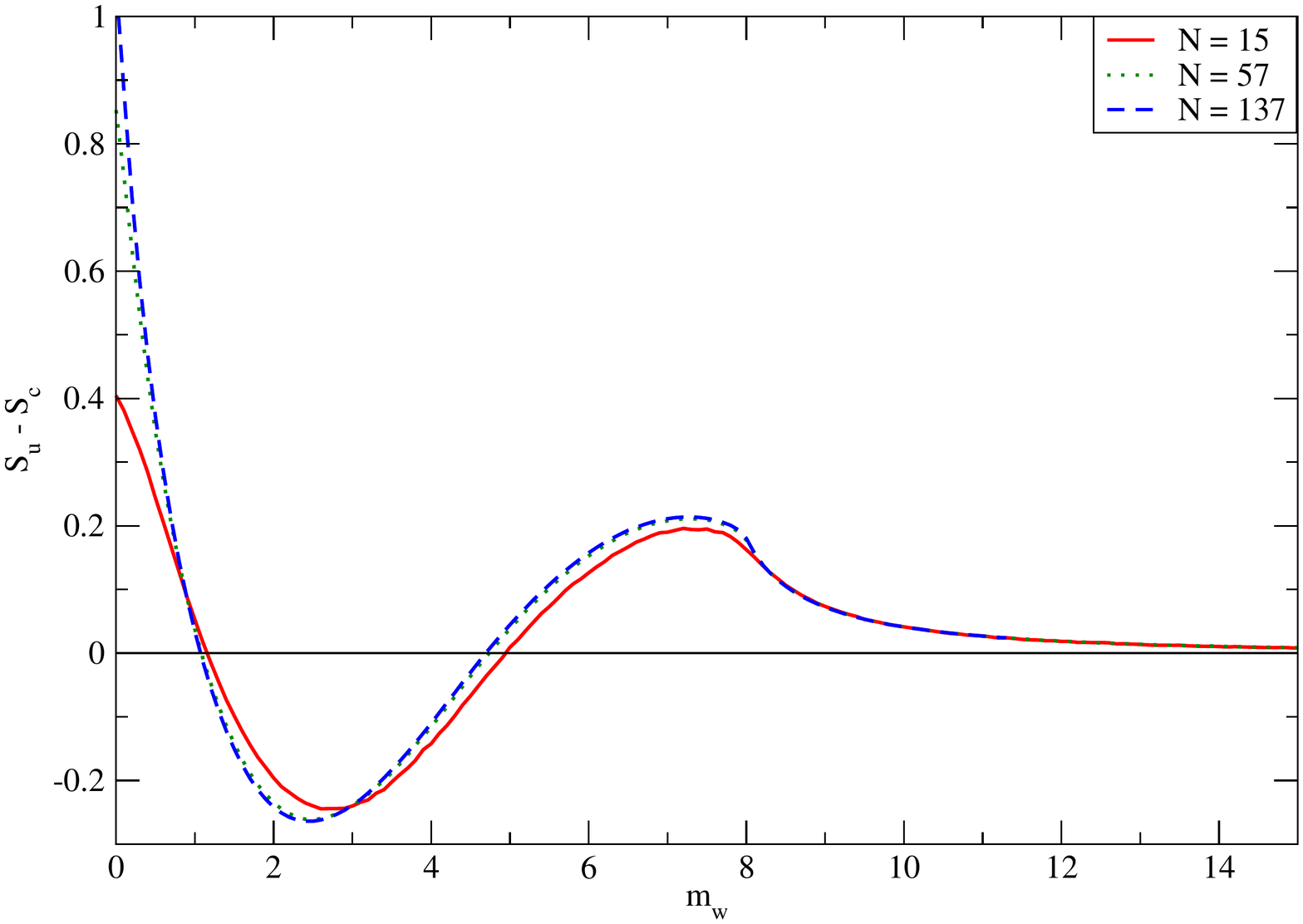}
    \caption{The difference between actions with correlated $S_c$ and
      uncorrelated eigenvalues to each direction.\label{act}}
  \end{center}
\end{figure}

It is possible that center symmetry is broken in a subtle manner
in the range $3.0< m_w < 10.0$. For example, the eigenvalues of
the individual Polyakov loop operator might be uniformly distributed
but they might show correlations in different directions\footnote{This is the problem with quenched
  Equchi-Kawai model \cite{Bringoltz:2008av}.}. In order to have
a correct sum over all {\sl momenta} as one would have in a infinite
lattice,
we need to ensure that the traces of Polyakov loops vanish and there are
no correlations between different directions
on the single site model.
Let us assume that
the eigenvalues are uniformly distributed and choose
$\frac{2\pi j}{N}$, $j=1,\cdots, N$ as the $N$ eigenvalues.
Let $\pi_j$, $j=1\cdots, N$ denote a permutation of $j=1\cdots, N$.
We compute the correlated action, $S_c$, with 
$\theta^j_\mu=\frac{2\pi j}{N}$ and compare it to
the uncorrelated action, $S_u$, with
$\theta^j_\mu=\frac{2\pi \pi^\mu_j}{N}$ where $\pi^\mu$ are different
permutations for different $\mu$.

\begin{figure}
    \begin{center}
    \includegraphics[width=1.0\textwidth]{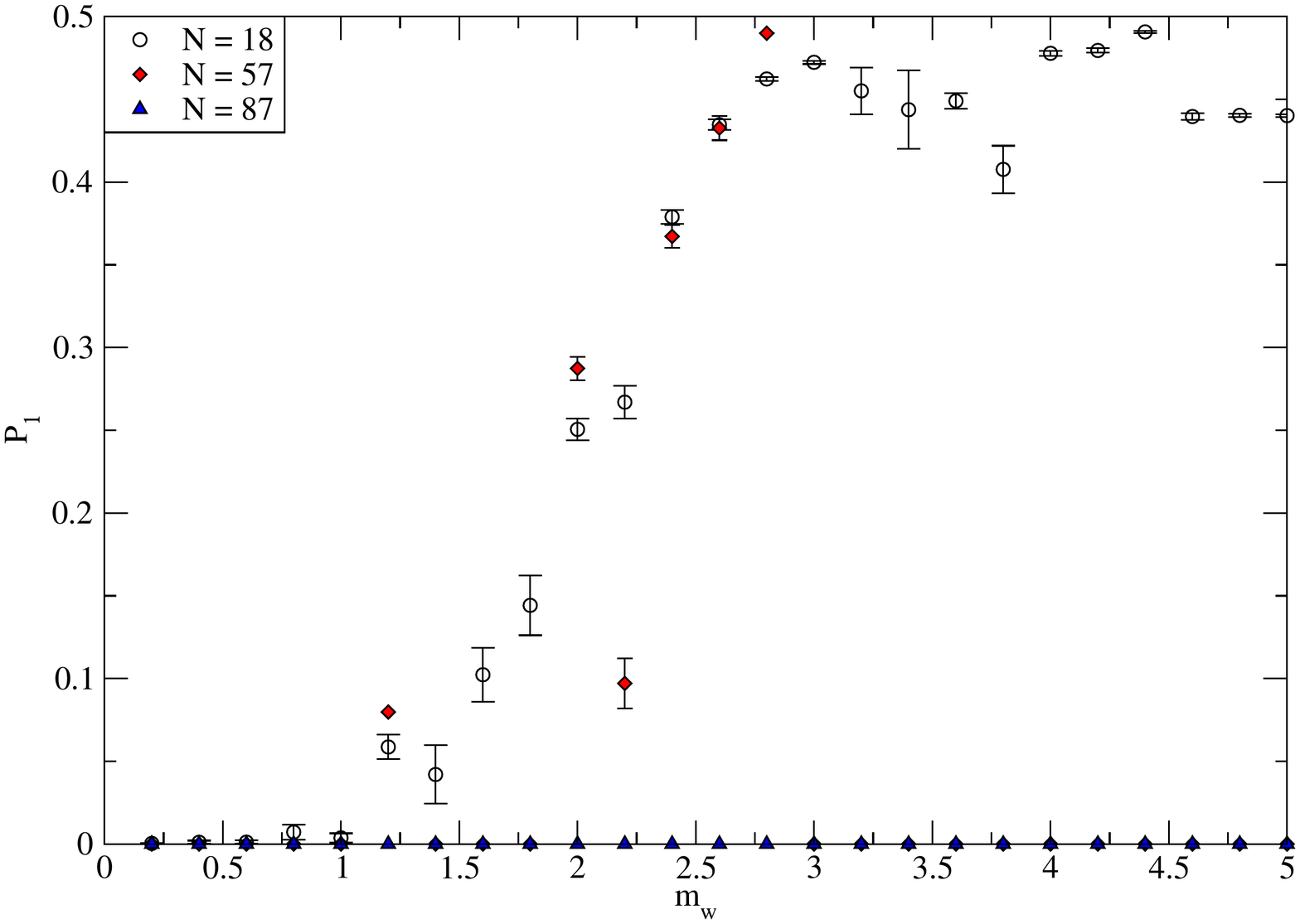}
    \caption{Plot of $P_\mu$ as a function of $m_w$ with massive quarks.\label{symm_mass}}
  \end{center}
\end{figure}

Fig.~\ref{act} shows the difference, $S_u-S_c$,  as a function of Wilson mass $m_w$ with different $N$. The value for $S_u$ is obtained by averaging over
several different random permutations but the fluctuations get smaller
as $N$ increases and it is sufficient to consider just one random
permutation as $N\to\infty$.
The uncorrelated minimum is preferred when
$1<m_w<5$. There is again virtually no dependence on $N$. This
combined with the restriction of center symmetry restoration gives 
\be 3<m_w<5 \label{allmw}\ee 
as the range for Wilson mass.  

One might wonder why the region of allowed $m_w$ does not include
zero.
In a typical free field analysis of overlap fermions~\cite{Narayanan:1994gw,Narayanan:1993sk}, 
one shows that it correctly represents a single Dirac flavor in a region around
zero momentum as long as $0< m_w < 2$. Momentum in our case
is replaced by $(\theta_\mu^i-\theta_\mu^j)$ and we want to
cover the whole range of allowed momenta. If this does not occur,
we will not have proper reduction or a correct realization of the
center symmetric phase. Because the range of allowed momenta (volume of
the Brillouin zone) in the
conventional free field analysis increases as $m_w$ increases,
we see why $m_w$ close to zero is not appropriate. Since we do
not have a concept of doublers on a single site lattice, we do not
require $m_w < 2$~\cite{Hietanen:2009ex}. Therefore, to reach momenta
close to $\pi$ and
have proper sampling of all momenta as per the infinite
lattice, we find a range of allowed $m_w$ than includes $m_w>2$.
One can also understand why $m_w$ cannot be arbitrarily large since
we would be approaching the limit of na\"ive fermions which does
not have a center symmetric phase on a single site lattice~\cite{Hietanen:2009ex}.

Once fermions have a non-zero mass, the fermionic contribution to
(\ref{pertact}) is always finite. Then the minimum of the perturbative
action is dominated by the pure gauge part and occurs when all the
eigenvalues are the same. 
The effect of finite $N$ is 
is demonstrated  in Fig.~\ref{symm_mass}
for $\mu=0.1$. We have only plotted the component $P_1$, since it
determines the center symmetry breaking point.
The symmetry breaking is evident as $N\to\infty$.

\begin{figure}
  \begin{center}
    \includegraphics[width=1.0\textwidth]{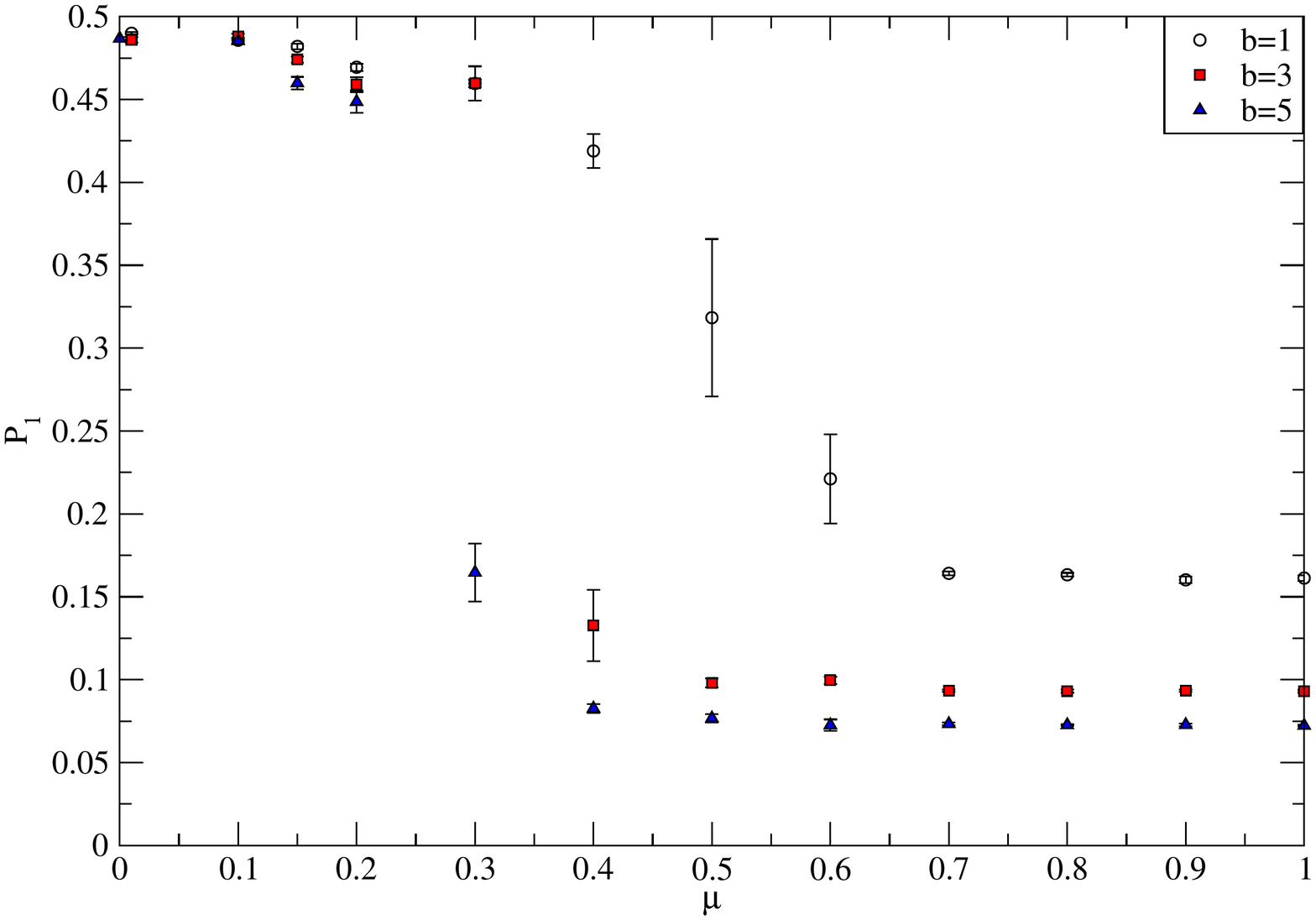}
  \caption{Plot of $P_1$ as a function of mass for three different $b$ with $N=15$ \label{csymmmN15}}
  \end{center}
\end{figure}

For the actual lattice simulation, we used HMC-algorithm described
in~\cite{Hietanen:2009ex}. All the simulations were performed with
$f=0.5$, Wilson mass $m_w=5$,~\footnote{This value is slightly high,
  since it is on the high end of (\ref{allmw}).  This
  is because the argument presented with regard to Fig.~\ref{act} was
  realized
after we obtained the numerical results presented in this section.}
and they
consist of about 100 independent measurements. Thermalization is fast
and requires only about ten iterations. Most of the simulations were
performed with $N=15$, but to study $1/N$ effects we did also
simulations with $N=11$ and $N=18$.  The purpose of the simulations
are to find out the critical mass $\mu_c$ for center symmetry breaking
as a function of $N$ and $b$.

\begin{figure}
  \begin{center}
    \includegraphics[width=1.0\textwidth]{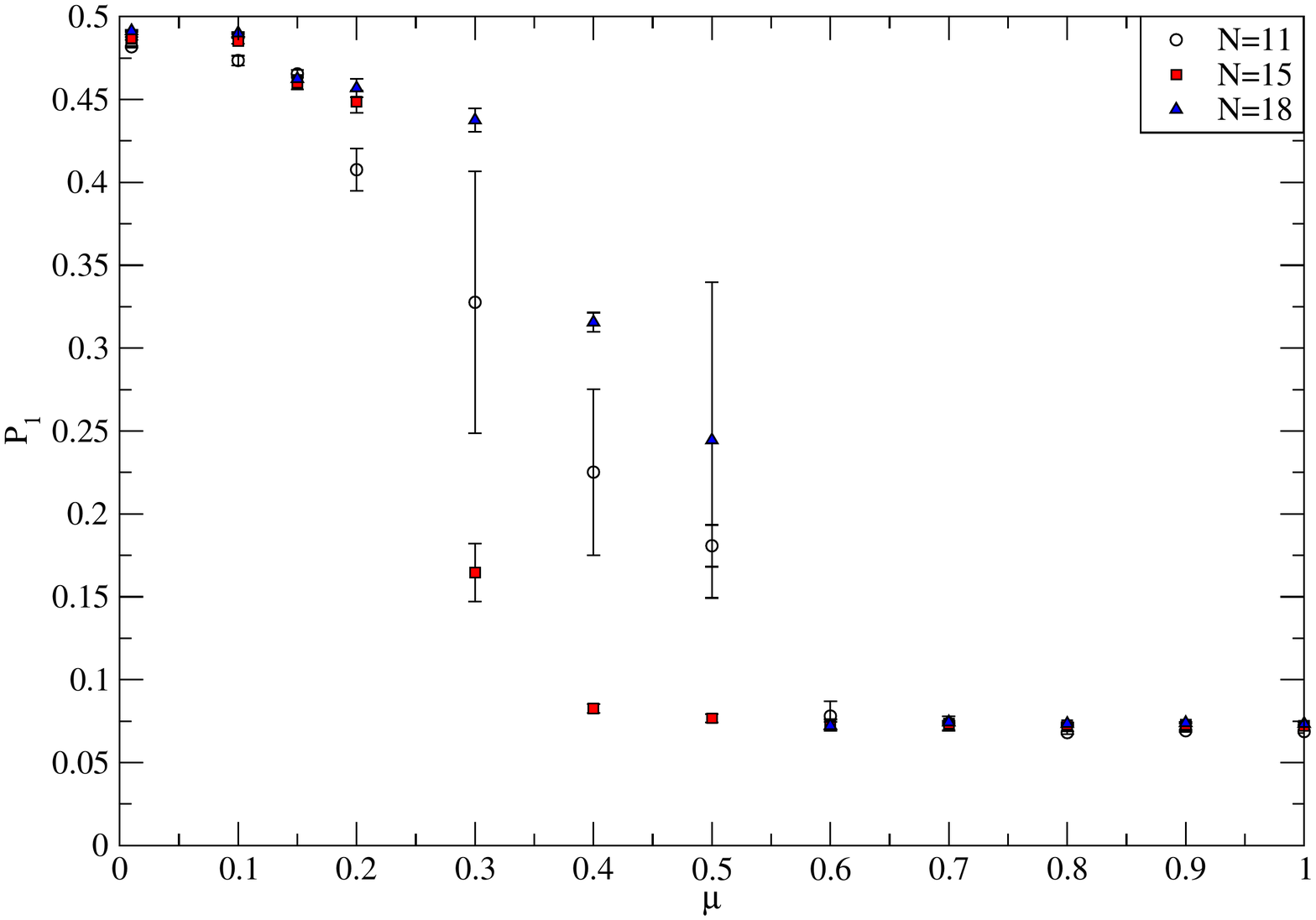}
    \caption{Plot of $P_1$ as a function of mass for three different $N$ with $b=5$ \label{csymmN}}
  \end{center}
\end{figure}

In Fig.~\ref{csymmmN15} we have plotted $P_1$ as a function of mass
with $N=15$ for $b=1,3$, and $5$. 
The data shows that $\mu_c(b)$ does decrease with increasing $b$ but 
the decrease is clearly slower than scaling would dictate. The range
of
center symmetry breaking is between $0.1$ and $0.3$ for $b$ in the
range $[1,5]$. Ignoring wave function renormalization, the dominant
part of the scaling dictates that we need to keep
$
\mu e^{\frac{8\pi^2}{3}b}
$
fixed as we take $b\to\infty$  in order to take the continuum limit at
a fixed physical mass. Our data for $\mu_c(b)$, therefore,
clearly indicates that we can take the continuum limit of a massive
adjoint fermion coupled to a large $N$ gauge field without any restriction
on its physical mass.

To understand the effects of finite $N$,
we performed simulations with 
$b=5$ also at $N=11$ and $N=18$. The $1/N$ effects are rather small
except in the region of the phase transitions as can be seen in Fig.~\ref{csymmN}. The critical value at $b=5$ is about $0.1$. 

We acknowledge discussions with Herbert Neuberger and Stephen Sharpe.
The authors acknowledge partial support by the NSF under grant number
PHY-0854744.  A.H. also acknowledges partial support by the
U.S. DOE grant under Contract DE-FG02-01ER41172.

\end{document}